# Pseudo-CTs from T1-weighted MRI for planning of low-intensity transcranial focused ultrasound neuromodulation: an open-source tool

Siti Nurbaya Yaakub[1,2], Tristan A. White[1], Eric Kerfoot[3], Lennart Verhagen[4], Alexander Hammers[2], Elsa F. Fouragnan[1]

[1]Brain Research & Imaging Centre, School of Psychology, University of Plymouth, Plymouth, United Kingdom

[2]King's College London & Guy's and St Thomas' PET Centre, School of Biomedical Engineering & Imaging Sciences, King's College London, London, United Kingdom

[3]Department of Biomedical Engineering, School of Biomedical Engineering & Imaging Sciences, King's College London, London, United Kingdom

[4]Donders Institute for Brain, Cognition and Behaviour, Radboud University, Nijmegen, the Netherlands

Correspondence:

Siti Nurbaya Yaakub

Brain Research & Imaging Centre, 7 Derriford Road, PL6 8BU, Plymouth, United Kingdom

siti.yaakub@plymouth.ac.uk

**ABSTRACT**

**Background:** Individual skull models of bone density and geometry are important when planning the expected transcranial ultrasound acoustic field and estimating mechanical and thermal safety in low-intensity transcranial ultrasound stimulation (TUS) studies. Computed tomography (CT) images have typically been used to estimate skull acoustic properties. However, obtaining CT images in research participants may be prohibitive due to exposure to ionising radiation and limited access to CT scanners within research groups.

**Objective:** We present a validated open-source tool for researchers to obtain individual skull estimates from T1-weighted MR images, for use in acoustic simulations.

**Methods:** We refined a previously trained and validated 3D convolutional neural network (CNN) to generate 100 keV pseudo-CTs. The network was pretrained on 110 individuals and refined and tested on a database of 37 healthy control individuals. We compared simulations based on reference CTs to simulations based on our pseudo-CTs and binary skull masks, a common alternative in the absence of CT.

**Results:** Compared with reference CTs, our CNN produced pseudo-CTs with a mean absolute error of 109.8 ± 13.0 HU across the whole head and 319.3 ± 31.9 HU in the skull. In acoustic simulations, the focal pressure was statistically equivalent for simulations based on reference CT and pseudo-CT (0.48 ± 0.04 MPa and 0.50 ± 0.04 MPa respectively) but not for binary skull masks (0.28 ± 0.05 MPa).

**Conclusions:** We show that our network can produce pseudo-CT comparable to reference CTs in healthy individuals, and that these can be used in acoustic simulations.

**INTRODUCTION**

Low intensity focused transcranial ultrasound stimulation (TUS) is an emerging technique for non-invasive neuromodulation. A variety of brain regions can potentially be targeted with TUS ranging from superficial cortical regions, such as the primary motor (Legon et al., 2018) and visual cortices (Lee et al., 2016), to deep cortical and sub-cortical regions like the anterior cingulate cortex (Fouragnan et al., 2019) and amygdala (Folloni et al., 2019). TUS has very high spatial specificity compared with other neuromodulatory techniques like transcranial magnetic or transcranial electric stimulation (TMS/tES), with TUS foci in the order of millimetres (Blackmore et al., 2019) compared to centimetres in TMS and tES. Due to this high spatial specificity, precisely identifying the location of the TUS focus and its acoustic profile (Mueller et al., 2017, 2016) is very important. Because bone is a major obstacle to acoustic waves, simulations of ultrasound wave propagation through the skull via accurate models are an essential component in planning TUS experiments, in terms of safety, efficiency and accuracy, as they enable pressure and temperature profiles in the brain to be mapped.

The skull accounts for the bulk of transcranial ultrasound attenuation and aberration. Individual skull models of bone density and geometry are thus important when planning the expected acoustic field in TUS experiments. Computed tomography (CT) images are considered the gold standard for skull imaging. CT images have good contrast between bone and soft tissue, and CT Hounsfield units (HU) have been used to estimate skull acoustic properties in *ex vivo* experiments (Aubry et al., 2003). However, obtaining CT images in research participants may be prohibitive due to exposure to ionising radiation and limited access to CT scanners within research groups.

Magnetic resonance (MR)-based methods for skull imaging are an alternative to CT. Short echo time (TE) sequences such as ultra-short echo time (UTE) or zero echo time (ZTE) MR imaging have been developed to produce images of the skull. However, bone cannot be directly imaged with MR because of its short $T2^*$. The short TE MR sequences instead image water bound to minerals in the bone or free water within spaces or pores in the bone (Ma et al., 2020). MR-based methods for skull imaging cannot be straightforwardly scaled to CT HU, the units most commonly used to estimate acoustic properties of the skull in simulations.

Deep learning methods involving convolutional neural networks (CNNs) are emerging as a tool for translating MR to CT images. To date there are three studies that have evaluated CNN-based MR to CT image translation for use in transcranial ultrasound planning. Su and colleagues proposed a 2D U-Net to synthesise pseudo-CT from a dual echo UTE MR image, and showed that these could be used in the planning of MR-guided focused ultrasound intervention (Su et al., 2020). Using a 3D generative adversarial network (GAN), Koh and colleagues synthesised pseudo-CT from T1-weighted MR and showed its utility in TUS planning (Koh et al., 2021). However, these methods tend to be trained and tested on relatively small datasets (n = 41 and 15 respectively) and may not be able to capture the wide range of inter-individual variability in skull geometry and density, limiting their generalisability. More recently, Miscouridou and colleagues used a 2D U-Net trained on stacks of 2D transverse slices (to incorporate 3D information) to produce pseudo-CT from T1-weighted and ZTE MR images, and showed that these give comparable accuracy to CT in acoustic simulations (Miscouridou et al., 2022).

Despite the development of these machine learning models to generate pseudo-CT from MR images, uptake of these techniques among the growing community of TUS researchers remains low as these models are not easily implemented by the non-expert. As a result, many low-intensity TUS research studies either do not report transcranial acoustic simulations, apply a fixed percentage derating of free-water values, or base their simulations on binary skull models or an example CT from one individual. Our primary aim in this study is to produce a well-validated open-source tool for researchers to obtain individual skull estimates from T1-weighted MR images for each of their study participants to use in acoustic simulations to accelerate replicability and best practice.

We use a 3D residual U-Net to synthesise pseudo-CT from T1-weighted MR images. Our network was pre-trained on a large number of subjects (n = 110) in order to better capture inter-individual variability, and potentially more able to generalise to other datasets. Our pre-trained network has previously been validated against other machine learning methods (e.g. multi-atlas co-registration) for pseudo-CT synthesis in a positron emission tomography MR-based attenuation correction application (Yaakub et al., in preparation). We further refined and validated our network on an separate dataset (n = 37) with different CT and MR properties to the initial pre-trained network. We produced 100 keV pseudo-CTs for use in acoustic simulations for planning low-intensity TUS experiments and compared these to acoustic simulations from real CTs and using binary skull masks. Our pre-trained network is also made available for others to use as a starting point to refine the network weights in their own datasets.

## MATERIAL AND METHODS

### Data

Details of the larger training dataset and the network implementation and validation are given in (Yaakub et al., in preparation). Briefly, our training and testing data were pairs of 3D T1-weighted magnetisation-prepared rapid gradient-echo (MPRAGE) MR images acquired on a 3.0T Siemens Biograph mMR PET-MR and low-dose CT (140kVp; GE Discovery 710 PET/CT system) images acquired in 110 individuals (mean age ± SD = 34.0 ± 17.6 years; range = 9 – 79 years; 61 female) on the same day. The 110 images were from healthy control individuals and patients who were referred for a PET-CT scan at St Thomas' Hospital, London, UK.

We refined and validated our network on data from 37 healthy control individuals from the CERMEP-IDB-MRXFDG Database (Mérida et al., 2021) (mean age = 38.1 ± 11.4; 20 female). The database consisted of 3D T1-weighted MPRAGE MR images acquired on a 1.5T Siemens Sonata scanner (matrix size = 160 × 192 × 192, voxel size = 1.2 × 1.2 × 1.2 mm$^3$) and low-dose CT (100 keV) data acquired on a Siemens Biograph mCT64 (voxel size = 0.6 × 0.6 × 1.5 mm$^3$). We bias-corrected the T1-weighted MR data using N4ITK bias field correction (Tustison et al., 2010) and edited the scanner bed out of the CT images by creating a head mask from the CT images using Otsu thresholding, smoothing and dilation. We applied the same binary head mask to the T1-weighted MR data to reduce the effect of scanner noise outside the head. To match the images to our acoustic simulation grid size, we resampled the MR images to 1 mm$^3$ isotropic resolution (resulting matrix size = 191 × 229 × 229) and rigidly realigned and resampled the CT to the MR images using NiftyReg (Modat et al., 2014) with cubic spline interpolation.

**Network architecture, training, and validation**

*Network architecture & training*

Our 3D CNN (Figure 1) was based on a residual U-Net (Kerfoot et al., 2019) with sub-pixel upsampling convolutional layers (Shi et al., 2016) implemented in PyTorch with the MONAI framework (https://monai.io) (MONAI Consortium, 2022). Our network took as input a 3D patch of 32 contiguous slices of MR data in the transverse/axial plane (256 × 256 × 32 voxels) and produced the corresponding pseudo-CT patch. The network was trained with a dropout rate of 0.1.

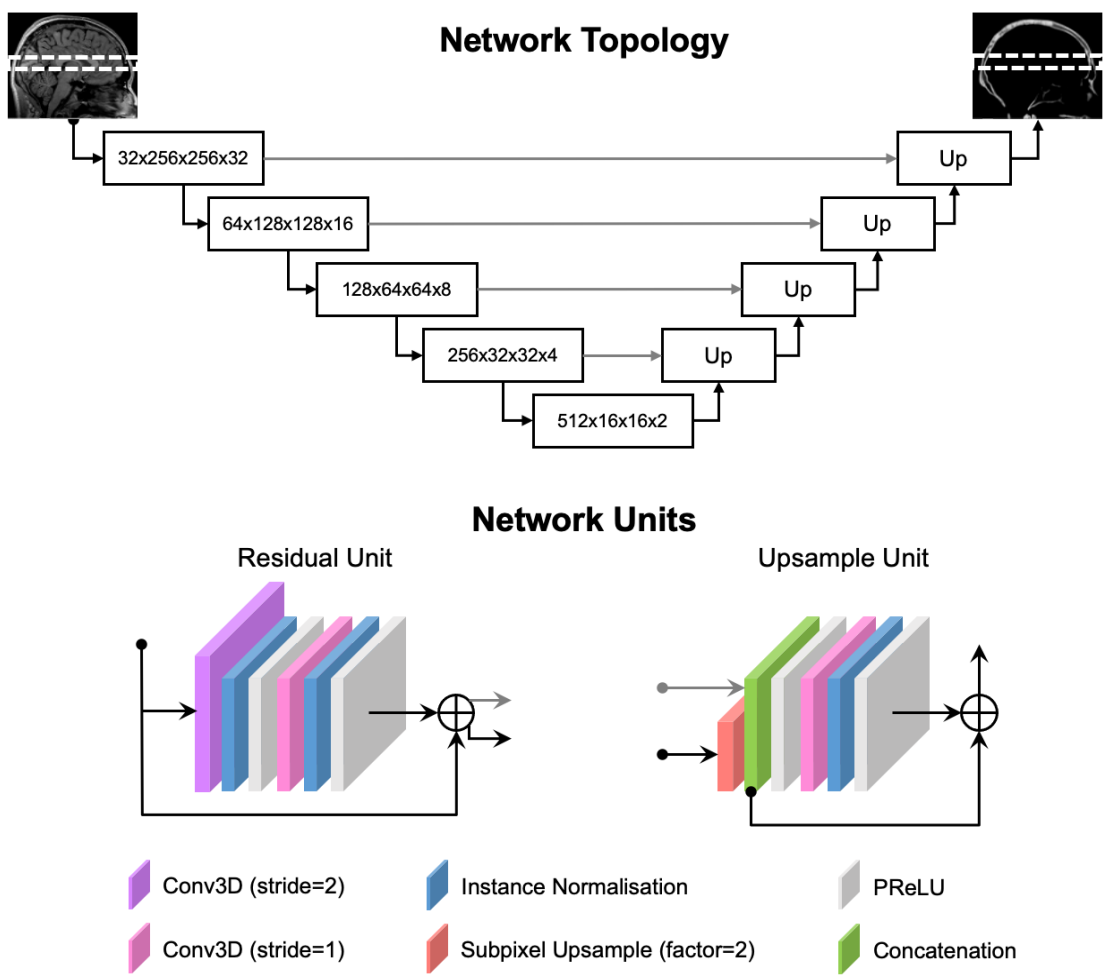

**Figure 1.** 3D residual U-Net with subpixel upsampling. Overall network topology (top) and details of network units (bottom). In the top panel, the number of channels and tensor size are shown in each layer of the encoding path. Conv3D: 3D convolutions; PReLU: Parametric Rectified Linear Unit.

We used the pre-trained weights from training the network on 110 individuals to further refine our network with a four-fold cross-validation method (~75% train; ~25% test/validation in each fold) in the 37 datasets from the CERMEP database. Input MR volumes were normalised using the default *HistogramNormalize* transform in MONAI and rescaled between -1 and 1, so that all MR images were in the same range. Reference CT images were thresholded at -1000 HU and intensity values above 2000 HU were clamped to reduce the influence of high-intensity voxels, typically found in the teeth or fillings in CT images, on the overall range of intensities in the CT images. MR and CT images were padded using the minimum value in each image (i.e. -1 for MR and -1000 for CT images) to obtain a matrix size of 256 × 256 × 256. Patches of size 256 × 256 × 32 were then randomly sampled for each image pair to use as input data to train the network. For each fold of the cross-validation, we initialised the network with the pre-trained weights and used the AdamW optimiser (Loshchilov and Hutter, 2019) with an initial learning rate of 0.003 and decay rate of 0.2. We trained the network on batches of 10 patches for 250 epochs before reducing the learning rate by a factor of 10 and the decay rate by a factor of two (i.e. learning rate = $1 \times 10^{-4}$ and decay = 0.1), and training for a further 500 epochs.

The full 3D pseudo-CT volumes were produced from T1-w MR images of each test subject by predicting patches with overlaps of 30 slices in the transverse/axial plane. We averaged voxel intensities in overlapping predicted slices.

*Network validation: accuracy of pseudo-CTs*

The accuracy of the pseudo-CTs synthesised by our method was assessed against reference CTs using the mean absolute error (MAE) metric (Burgos et al., 2014) within a head mask ($MAE_{head}$). Additionally, using a threshold of 300 HU to define a skull mask, we quantified the MAE between reference CT and pseudo-CT within the skull ($MAE_{skull}$) and quantified the amount of overlap between the skull masks derived from the reference CT and pseudo-CT using the Jaccard coefficient (Jaccard, 1901; intersection over union) and the Dice similarity coefficient (Dice, 1945).

**Validation of pseudo-CTs in acoustic simulations**

*Target and neuronavigation*

We chose the dorsal anterior cingulate cortex (dACC), a deep cortical region, as our target region for TUS acoustic simulations. Neuronavigation was performed with the Brainsight TMS software (Rogue Research Inc., Montréal, Québec, Canada) using the T1-weighted MR images as the head model. Our simulated transducer was based on the NeuroFUS TPO & CTX-500-4 transducer (Brainbox Ltd., Cardiff, UK), which consisted of a four-element annular transducer with a central frequency of 500 kHz. We allowed for 11 mm of space between our simulated transducer face and the scalp (to account for the use of a gel pad or water bladder for coupling) and optimised the trajectory so that it was perpendicular to the skin surface in each subject. All neuronavigation procedures were performed in native space. We normalised images to the Montreal Neurological Institute (MNI) 1mm$^3$ brain template using NiftyReg and used the inverse transformation parameters to initialise the dACC target at MNI coordinates: x = -5, y = 24, z = 30. Where necessary, the target was then adjusted for each participant so that it was centred in the grey matter. All neuronavigation and simulations were done in native space, the transform to MNI space was only for initialising the target location.

*Acoustic simulation*

We used the k-Wave Toolbox (version 1.3) (Treeby and Cox, 2010) for our simulations performed in MATLAB (R2020b, MathWorks, Inc.). Our code for performing acoustic simulations is available on GitHub: https://github.com/sitiny/BRIC_TUS_Simulation_Tools. Our simulation grid size was 256 × 256 × 256 points with a grid spacing of 1 mm, corresponding to three points per wavelength (PPW) at 500 kHz. We set the maximum time step using the *checkStability* function in k-Wave, which resulted in a Courant-Friedrichs-Lewy (CFL) number of 0.202 and ran the simulation for 200 μs.

The skull was approximated from CT images by thresholding at 300 HU and clamping values above 2000 HU as previously stated. CT HU were then linearly mapped to acoustic properties with the following equations for density ($\rho$), speed of sound ($c$) and absorption coefficient ($\alpha$) (Marsac et al., 2017; Mueller et al., 2017):

$$\rho_{skull} = \rho_{water} + (\rho_{bone} - \rho_{water})\frac{HU - HU_{min}}{HU_{max} - HU_{min}} \qquad (1)$$

$$c_{skull} = c_{water} + (c_{bone} - c_{water}) \frac{\rho_{skull} - \rho_{water}}{\rho_{bone} - \rho_{water}} \qquad (2)$$

$$\alpha_{skull} = \alpha_{bone_{min}} + (\alpha_{bone_{max}} - \alpha_{bone_{min}}) \left(1 - \frac{HU - HU_{min}}{HU_{max} - HU_{min}}\right)^{0.5} \qquad (3)$$

where $HU_{min}$ and $HU_{max}$ were the minimum and maximum HU in the skull. The acoustic parameters we used in the simulations are given in Table 1, with values for the density and sound speed from (Bancel et al., 2021) and absorption coefficients at 500 kHz from (Fry and Barger, 1978). All other media outside the skull were treated as homogeneous with acoustic properties of water.

**Table 1.** Acoustic parameters used in simulations (Bancel et al., 2021; Fry and Barger, 1978).

| Density (kg/m³) | Sound speed (m/s) | Absorption (dB/MHz cm) |
|---|---|---|
| $\rho_{water} = 1000$ | $c_{water} = 1500$ | $\alpha_{water} = 0$ |
| $\rho_{bone} = 1900$ | $c_{skull} = 3100$ | $\alpha_{bone_{min}} = 4$; $\alpha_{bone_{max}} = 8.7$ |

We performed acoustic simulations to match a spatial peak pulse averaged intensity ($I_{SPPA}$) of 20.00 W/cm² in free field (corresponding to a maximum pressure of 0.775 MPa and mechanical index (MI) of 1.095). The four-element transducer was modelled using the kWaveArray tools (alpha 0.3) from the k-Wave toolbox. We simulated the acoustic field for a range of focal depths in free field to obtain the corresponding combinations of the relative phases and magnitudes of each element while keeping the free-field $I_{SPPA}$ constant at 20 W/cm² (Supplementary Table 1). We then performed the simulations using each individual's CT to estimate the skull with the combination of phase and magnitude (from the free field simulations) that matched the focal depth of their dACC, based on the neuronavigation.

We repeated the acoustic simulations using our pseudo-CTs to estimate the skull (instead of the reference CTs), keeping the simulation parameters the same as the simulations based on reference CTs. We reported the maximum focal pressure, MI, and $I_{SPPA}$ and the difference in focal position and -6 dB focal volume compared to reference CT-based simulations. We tested the equivalence of maximum focal pressure between simulations based on the reference CT skulls and pseudo-CT skulls using the two one-sided test of equivalence for paired samples (TOST-P) procedure. The smallest effect size of interest was 10% for focal pressure, and was based on the median difference values in acoustic simulation software intercomparison benchmarks (Aubry et al., 2022).

A common method used by research groups for acoustic simulations is to use a binary skull model. To show the added benefit of using a pseudo-CT skull model, we additionally performed simulations using binary skull models (derived from reference CTs by thresholding at 300 HU) and compared the simulation output to simulations based on reference CT skulls as above.

**Example application to novel data**

To test the generalizability of our method, we applied the CNN to novel datasets that were acquired on different scanners and sites to the CERMEP database. We re-trained our network using all 37 datasets available from the CERMEP database, with the same training scheme as in the four-fold cross-validation. The data we applied our trained network to were: 1) five T1-weighted MR images (MPRAGE sequence acquired in sagittal plane, 2100 ms repetition time (TR), 2.26 ms echo time (TE), 900 ms inversion time (TI), 8° flip angle (FA), GRAPPA acceleration factor of 2, and 1 mm$^3$ voxel size) acquired on a 3T Siemens MAGNETOM Prisma scanner (Siemens Healthcare GmbH, Erlangen, Germany) at the Brain Research & Imaging Centre (BRIC) in Plymouth, United Kingdom, as part of a larger TUS study; and 2) five T1-weighted MR images (MPRAGE sequence acquired in sagittal plane, 2700 ms TR, 3.69 ms TE, 1090 ms TI, 9° FA, GRAPPA acceleration factor 2, water excitation fat suppression and $0.9 \times 0.9 \times 0.9$ mm$^3$ voxel size) acquired on a 3T Siemens MAGNETOM Skyra scanner (Siemens Healthcare GmbH, Erlangen, Germany) at the Donders Institute in Nijmegen, The Netherlands, as part of another larger TUS study.

The T1-weighted MR images were first pre-processed in the same way as the input training data to the network: we applied bias correction using N4ITK bias field correction. Since there was no reference CT to derive a head mask from, we created a head mask derived from Otsu thresholding, smoothing and dilation of the MR images directly. We quantified the mean, maximum and minimum intensities in the pseudo-CT produced and visually assessed how well the pseudo-CT overlapped with skull on the MR image.

# RESULTS

## Accuracy of pseudo-CTs obtained with our method

When compared to the reference CTs, our CNN produced pseudo-CT images with $MAE_{head}$ of $109.8 \pm 13.0$ HU, and $MAE_{skull}$ of $319.3 \pm 31.9$ HU. The Jaccard coefficient of overlap between reference CTs and pseudo-CTs within the skull was $0.70 \pm 0.04$ (Dice coefficient = $0.82 \pm 0.03$). Table 2 summarises the MAE, Jaccard and Dice coefficients for each fold of the cross-validation.

**Table 2.** Accuracy of pseudo-CTs compared with reference CTs. $MAE_{head}$ and $MAE_{skull}$ denote the mean absolute error between reference CT and pseudo-CT within the head and skull respectively. Values are mean ± standard deviation.

| Fold | $MAE_{head}$ [HU] | $MAE_{skull}$ [HU] | Jaccard | Dice |
|---|---|---|---|---|
| 1 | 113.2 ± 12.6 | 324.3 ± 31.7 | 0.70 ± 0.04 | 0.83 ± 0.03 |
| 2 | 105.5 ± 5.3 | 311.2 ± 21.8 | 0.70 ± 0.04 | 0.82 ± 0.03 |
| 3 | 116.1 ± 15.7 | 330.7 ± 36.6 | 0.69 ± 0.04 | 0.82 ± 0.03 |
| 4 | 103.9 ± 14.2 | 310.5 ± 37.6 | 0.71 ± 0.04 | 0.83 ± 0.02 |
| Mean | 109.8 ± 13.0 | 319.3 ± 31.9 | 0.70 ± 0.04 | 0.82 ± 0.03 |

Figure 2 shows the T1-weighted MR image, reference CT, pseudo-CT and difference image (reference CT–pseudo-CT) for the individual from the CERMEP dataset showing median performance based on $MAE_{head}$ ($MAE_{head}$ = 108.4 HU; $MAE_{skull}$ = 333.3 HU). The best and worst performing individuals (i.e. with lowest and highest $MAE_{head}$ respectively) are shown in Supplementary Figure 1.

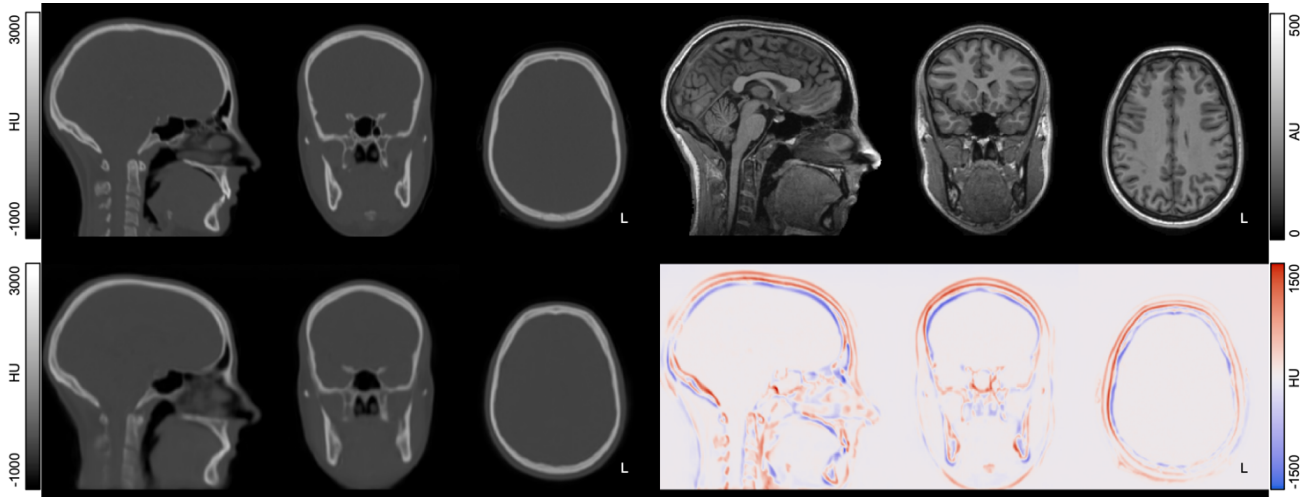

**Figure 2.** Reference CT (top left) and corresponding T1-weighted MR image (top right) for a representative individual (median MAE$_{head}$) shown with the pseudo-CT (bottom left) generated using our method and the difference image (reference CT–pseudo-CT; bottom right). L denotes left side of head. Image cross-sections are shown at x = 96, y = 132, z = 167 in the individual's native coordinate space.

**Validation of pseudo-CTs for acoustic simulations**

*Target and neuronavigation*

The dACC was identified for each participant in the database. The mean distance between the transducer face and the dACC target was 52.97 ± 2.83 mm, with a minimum focal distance of 48 mm and a maximum of 60 mm.

*Acoustic simulations*

Table 3 summarises the acoustic simulation metrics at the focus for skull models based on CT, pseudo-CT, and binary skull masks. The TOST-P procedure revealed that the mean focal pressure obtained from pseudo-CT-based simulations was statistically equivalent to reference CT-based simulations for the equivalence bounds of 10% of mean focal pressure [-0.048, 0.048] ($t_{36}$ = -3.69, p = 0.0004). The mean focal pressure obtained from binary skull-based simulations were not statistically equivalent to reference CT-based simulations ($t_{36}$ = -34.3, p = 1).

**Table 3.** Simulated metrics at the acoustic focus for skull models based on reference CT, pseudo-CT and binary skull. Values are given as mean ± SD, and the minimum – maximum value range across the 37 individuals. Differences in focal position and focal volume are given with respect to reference CT simulations. MI: mechanical index, $I_{SPPA}$: spatial peak pulse averaged intensity.

| Skull model | Attenuation (%) | Maximum Pressure [MPa] | MI | $I_{SPPA}$ [W/cm$^2$] | Difference in focal position [mm] | Absolute difference in focal volume [%] |
|---|---|---|---|---|---|---|
| Reference CT | 61.1 ± 6.7%<br>46.0 – 79.4% | 0.48 ± 0.04<br>(0.35 – 0.57) | 0.68 ± 0.06<br>(0.50 – 0.81) | 7.79 ± 1.34<br>(4.11 – 10.80) | - | - |
| Pseudo-CT | 57.6 ± 5.9%<br>48.3 – 70.8% | 0.50 ± 0.04<br>(0.42 – 0.56) | 0.71 ± 0.05<br>(0.59 – 0.79) | 8.48 ± 1.19<br>(5.85 – 10.35) | 2.6 ± 1.8<br>(0.0 – 8.6) | 17.3 ± 12.9<br>(0.0 – 43.2) |
| Binary skull | 87.5 ± 2.5%<br>81.3 – 92.8% | 0.28 ± 0.05<br>(0.22 – 0.51) | 0.50 ± 0.07<br>(0.40 – 0.73) | 2.53 ± 0.48<br>(1.62 – 3.75) | 2.6 ± 1.8<br>(0.0 – 7.1) | 25.2 ± 16.5<br>(0.8 – 80.2) |

Figure 3 shows the simulated TUS pressure field overlaid on the MR for simulations based on CT, pseudo-CT and binary skulls in the same individual as in Figure 2 (i.e. with median MAE$_{head}$). Supplementary Figure 2 shows the acoustic simulation based on the three different skull models for the best and worst performing individuals based on MAE$_{head}$.

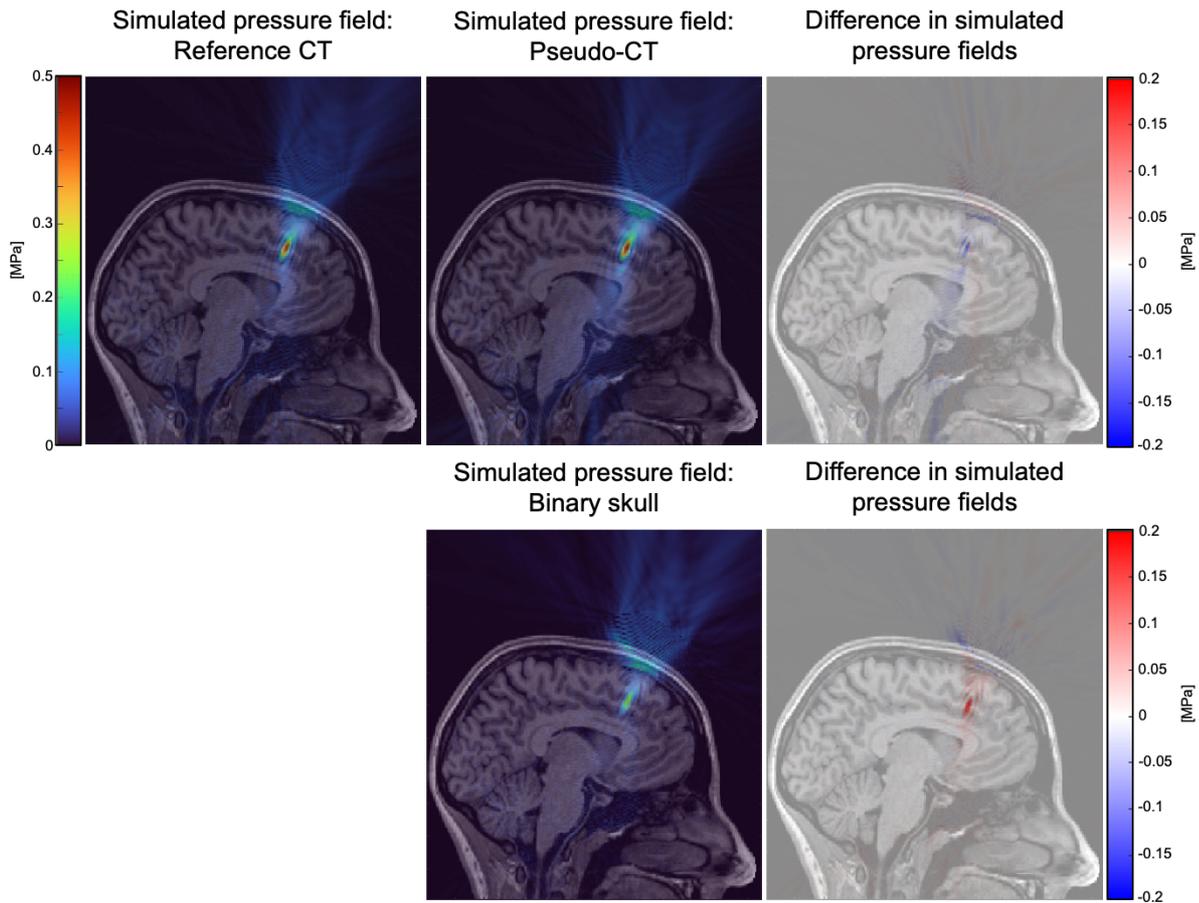

**Figure 3.** Simulated TUS pressure field for simulations based on reference CT (top left) overlaid on T1-weighted MR image for a representative individual (median $MAE_{head}$). Simulated TUS pressure field for simulations based on pseudo-CT and binary skull are shown in the middle column, and their differences with respect to the reference CT simulation (reference CT – pseudo-CT, or binary skull) are shown on the right. Sagittal cross-sections are shown at x = 90 in the individual's native coordinate space.

**Example application to novel data**

Figure 4 shows the MR and corresponding pseudo-CT produced by our network for the images acquired at BRIC and Donders. Visual inspection showed that the network performed better in the images from BRIC. High intensities were incorrectly assigned to the region around the head in the Donders images due to the intensities from fat matching that of bone in these images because they were acquired with fat suppression.

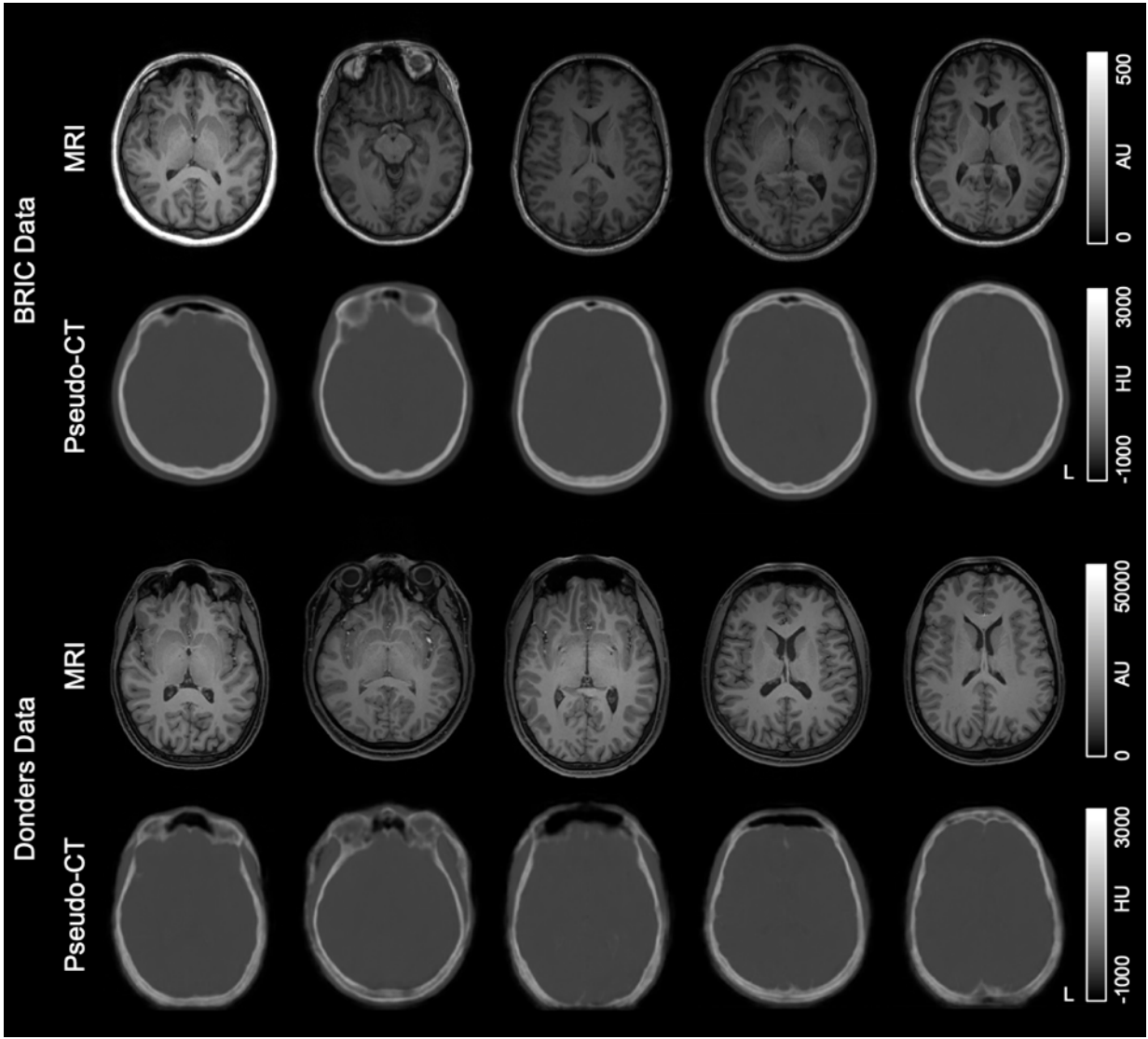

**Figure 4.** Application of network to novel data. Example axial slices of MR images and the corresponding pseudo-CT images produced by the network in novel data acquired at BRIC (top) and Donders (bottom). L denotes left side of head. Axial plane shown at z = 140 for BRIC data and z = 170 for Donders data in native space.

## DISCUSSION

We show that our 3D CNN method can be used to synthesise accurate 100keV pseudo-CT images from T1-weighted MR images, and that these can be used to estimate the skull in TUS simulations to obtain similar results as when using CT images to estimate the skull. This study highlights the importance of obtaining subject-specific acoustic simulations since individual variations in skull density and geometry can influence the extent of attenuation and focusing of the ultrasound beam. We provide an open-source CNN-based method to obtain pseudo-CT from T1-weighted images (https://github.com/sitiny/mr-to-pct). Users can also use our pre-trained network weights as a starting point to further train the CNN on their own data. This provides researchers with a useful resource for the design and analysis of TUS studies.

Our network generates pseudo-CT with MAEs ($MAE_{head}$ = 109.8 ± 13.0, $MAE_{skull}$ = 319.3 ± 31.9) comparable to other CNN-based methods for generating pseudo-CT (c.f. (Su et al., 2020): $MAE_{head}$ (from UTE MR) = 104.57 ± 21.33 HU; (Koh et al., 2021): $MAE_{head}$ (from T1-weighted MR) = 85.72 ± 9.50 HU; (Miscouridou et al., 2022): $MAE_{head}$ (from T1-weighted MR) = 133 ± 46, $MAE_{head}$ (from ZTE MR) = 83 ± 26. Although our network performance is similar to others, there might be room for further improvements. Indeed, our goal was to provide a validated method and open-source tools for others to replicate or improve on. For example, pseudo-CT generation tends to work better from ZTE MR images compared to T1-weighted MR images, but publicly accessible databases of these, paired with reference CT images do not exist. Furthermore, most research groups will also have a standard T1-weighted MR sequence available on their MR scanners while ZTE MR sequences tend to be more specialised.

We also show that our pseudo-CTs are suitable for use in acoustic simulations, producing peak pressure values that are statistically equivalent to CT-based simulations. Based on the simulations, binary skull models may be sufficient for estimating the location of the TUS focus, but should not be relied on to estimate the amount of transcranial pressure of the TUS beam in the brain. This is expected because in our binary skull model, we assigned cortical bone values to the whole skull, thus the estimated intracranial pressure was much lower (i.e. higher estimated attenuation) than in simulations based on CT skulls. Simulations based on binary skull models cannot be used for assessing safety indices like MI and $I_{SPPA}$. It should be noted that refined skull models can be derived from further segmentation of skull into trabecular and cortical bone segments, thus making it possible to assign different acoustic properties to the different bone segments. However, this approach relies on precise localisation and separation of trabecular and cortical bone, which is difficult with T1-weighted MR images alone.

We show that our network also works better in novel data (i.e. acquired independently of the 1.5T MR images in the CERMEP database that the network was trained on) when the T1-weighted images are acquired with a sequence similar to the one originally used during training. The pseudo-CTs produced from T1-weighted MRs acquired at BRIC were qualitatively better than from the T1-weighted MRs acquired at the Donders Institute. This could be because the BRIC data were acquired without fat suppression (like the original CERMEP data), while the Donders data were acquired with fat suppression. We would recommend that users apply the network on T1-weighted MR images acquired with a sequence close to the one used for training where possible, or do transfer learning on their own datasets using our pre-trained weights.

The study was limited by the resolution and availability of training data. Higher resolution data would enable the acoustic simulations to be performed with a smaller grid spacing for higher simulation accuracy. Other MR sequences that have better separation between bone and cerebrospinal fluid (e.g. T2-weighted MR) or better separation between trabecular and cortical bone (e.g. UTE or ZTE MR sequences) may be of benefit to increase the accuracy of the pseudo-CT (Miscouridou et al., 2022), however, we chose to use T1-weighted MRs as these are the most readily available anatomical research sequence. The acquisition x-ray tube energy and reconstruction technique for CT images affects the conversion of CT HU to skull acoustic velocity (Webb et al., 2018) and may affect the acoustic simulation based on the pseudo-CT. This does not preclude other groups from using these data (if they have them available) to re-train and optimise the CNN using our pre-trained network weights as a starting point.

We did not do a comparison to existing methods for skull imaging (e.g. UTE sequences). Skull imaging methods would probably outperform our current CNN in the ability to distinguish between trabecular and cortical bone because of the limitations we listed above. However, methods to convert MR arbitrary units to Hounsfield units do not perform as well as CT-based methods in simulations (Miscouridou et al., 2022).

Another possible limitation was that we did not use the same head mask during training and during testing of novel images. Since we already had a working script to produce a good, subject-specific head mask from CT images, we decided to apply this to MR images during training. The head mask can also be derived through other methods including co-registering a template head mask to each individual, although these methods come with their own set of limitations such as reduced field of view of template heads and manual delineation of the template mask. The head mask, regardless of how it is derived, should not make a big difference to the performance of the network on new datasets, as long as it excludes voxels outside the head.

## ACKNOWLEDGEMENTS

E.F. and S.N.Y. are supported by the MRC grant MR/T023007/1. This work was supported by the UK Department of Health via the NIHR Comprehensive Biomedical Research Centre Award (COV-LT-0009) to Guy's and St Thomas' NHS Foundation Trust (in partnership with King's College London and King's College Hospital NHS Foundation Trust), and by the Wellcome Engineering and Physical Sciences Research Council Centre for Medical Engineering at King's College London (WT 203148/Z/16/Z).

CERMEP-IDB-MRXFDG Database (© Copyright CERMEP – Imagerie du vivant, www.cermep.fr and Hospices Civils de Lyon. All rights reserved.) provided jointly by CERMEP and Hospices Civils de Lyon (HCL) under free academic end user licence agreement.

## DATA AVAILABILITY

The code for generating pseudo-CT from T1-weighted MR images and for running the acoustic simulations as described in this work are available on GitHub: https://github.com/sitiny/mr-to-pct and https://github.com/sitiny/BRIC_TUS_Simulation_Tools. The pre-trained network weights and pseudo-CT produced by our method are available to download from: https://osf.io/e7sz9/.

**Supplementary Material**
**Pseudo-CTs from T1-weighted MRI for planning of low-intensity transcranial focused ultrasound neuromodulation: an open-source tool; Yaakub et al.**

**Supplementary Table 1.** Phase (in degrees; °) for each element of the transducer and source amplitude (Pa) for different focal distances (mm). The phase of each element was obtained from the NeuroFUS system by setting the focal distance. The source amplitude was obtained from simulations in water to match a spatial peak pulse averaged intensity ($I_{SPPA}$) of 20 W/cm$^2$ in free field.

| Focal Distance (mm) | Element 1 Phase (°) | Element 2 Phase (°) | Element 3 Phase (°) | Element 4 Phase (°) | Source Amplitude (Pa) |
|---|---|---|---|---|---|
| 60 | 0 | 325.7 | 291.3 | 257 | 50023.5 |
| 59 | 0 | 329.2 | 298.3 | 267.5 | 49263.0 |
| 58 | 0 | 332.8 | 305.5 | 278.3 | 48522.0 |
| 57 | 0 | 336.5 | 312.9 | 289.3 | 47794.0 |
| 56 | 0 | 340.2 | 320.4 | 300.5 | 47071.0 |
| 55 | 0 | 344 | 327.9 | 311.8 | 46379.0 |
| 54 | 0 | 347.7 | 335.5 | 323 | 45723.0 |
| 53 | 0 | 351.3 | 342.7 | 333.9 | 45123.0 |
| 52 | 0 | 354.9 | 349.8 | 344.6 | 44576.0 |
| 51 | 0 | 358.3 | 356.6 | 354.9 | 44070.0 |
| 50 | 0 | 1.5 | 3 | 4.6 | 43574.0 |
| 49 | 0 | 4.6 | 9.3 | 14 | 43152.2 |
| 48 | 0 | 7.9 | 15.8 | 23.7 | 42736.0 |

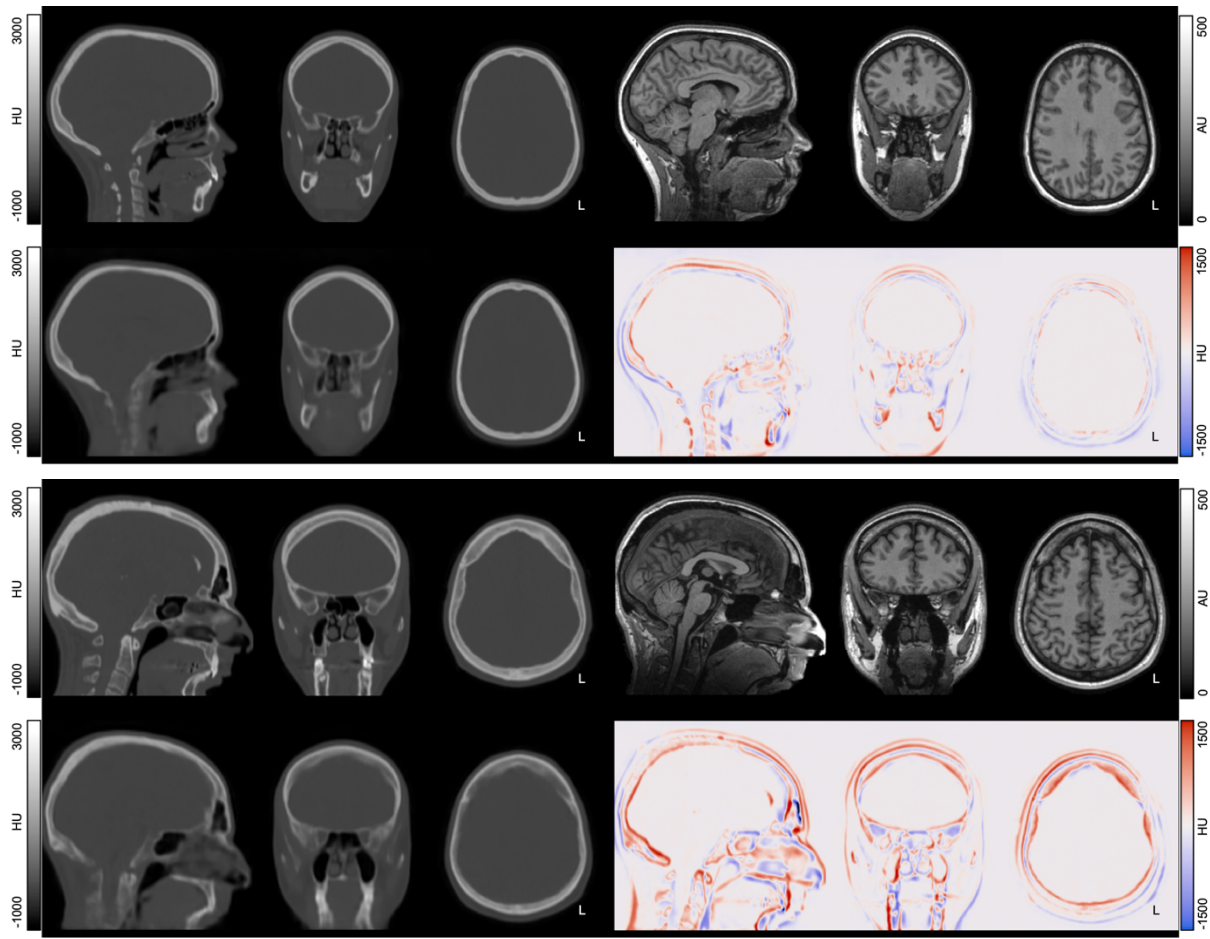

**Supplementary Figure 1.** Pseudo-CT vs. reference CT comparisons for best (top set of images) and worst (bottom set of images) performing individuals based on $MAE_{head}$. $MAE_{head}$ = 80.37 HU and 143.26 HU respectively. In each set of images, clockwise from top left: reference CT, corresponding T1-weighted MR image, difference between reference and pseudo-CT, pseudo-CT. L denotes left side of head. Image cross-sections are shown in native coordinate space x = 90, y = 143, z = 160 (top) and x = 95, y = 146, z = 167 (bottom).

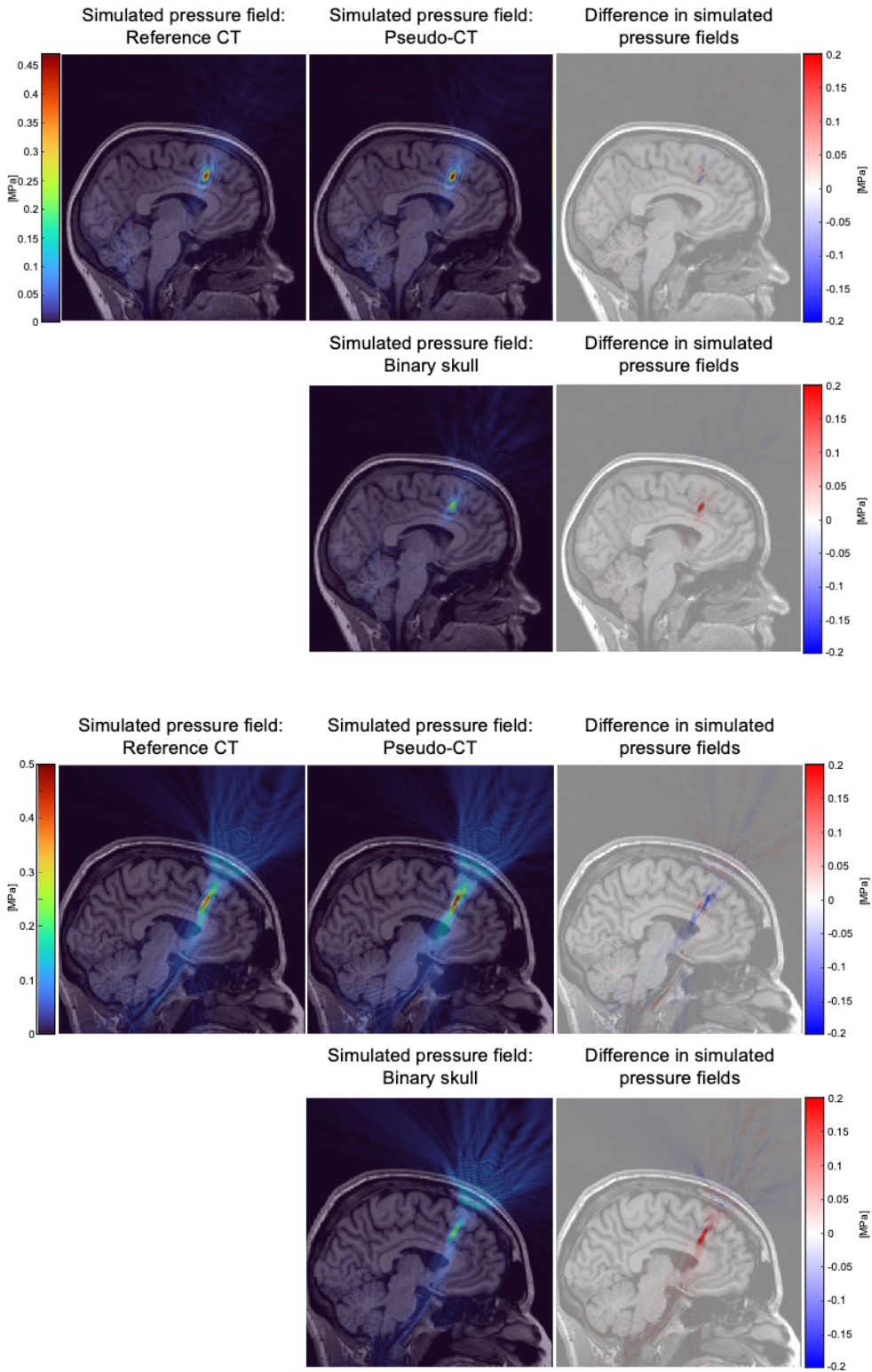

**Supplementary Figure 2.** Simulated TUS pressure field for simulations based on reference CT (top left) overlaid on T1-weighted MR image for best (top set of images) and worst (bottom set of images) performing individuals based on $MAE_{head}$. Simulated TUS pressure field for simulations based on pseudo-CT and binary skull are shown in the middle column, and their differences with respect to the reference CT simulation (reference CT – pseudo-CT, or binary skull) are shown on the right. Sagittal cross-sections are shown at x = 95 (top) and x = 90 (bottom) in the individual's native coordinate space.